\documentclass[journal=jpcafk,manuscript=article]{achemso}
\usepackage[version=3]{mhchem}
\usepackage{titlecaps}
\Addlcwords{the of into}
\usepackage{chemformula} 
\usepackage[T1]{fontenc} 
\usepackage{enumitem}
\usepackage{graphics}
\usepackage{bm}
\usepackage[normalem]{ulem}
\usepackage[symbol]{footmisc}

\Addlcwords{a an the of on at with for and in into from by without}

\usepackage{soul}
\usepackage{xr}
\makeatletter
\newcommand*{\addFileDependency}[1]{
 \typeout{(#1)}
 \@addtofilelist{#1}
 \IfFileExists{#1}{}{\typeout{No file #1.}}
}
\makeatother

\author{Juan Carlos San Vicente Veliz}
\affiliation{Department of Chemistry, University of Basel,
  Klingelbergstrasse 80, CH-4056 Basel, Switzerland}
  
\author{Julian Arnold}
\affiliation{Department of Physics, University of Basel,
  Klingelbergstrasse 82, CH-4056 Basel, Switzerland}
  
\author{Raymond J. Bemish} \affiliation{Air Force Research Laboratory,
  Space Vehicles Directorate, Kirtland AFB, New Mexico 87117, USA}

\author{Markus Meuwly}\email{m.meuwly@unibas.ch}
\affiliation{Department of Chemistry, University of Basel,
  Klingelbergstrasse 80, CH-4056 Basel, Switzerland}

\title[]{Combining Machine Learning and Spectroscopy to Model Reactive
  Atom + Diatom Collisions}

\abbreviations{}

\begin{document}

\date{\today}

\begin{abstract}
The prediction of product translational, vibrational, and rotational
energy distributions for arbitrary initial conditions for reactive
atom+diatom collisions is of considerable practical interest in
atmospheric re-entry. Due to the large number of accessible states,
determination of the necessary information from explicit
(quasi-classical or quantum) dynamics studies is impractical. Here, a
machine-learned (ML) model based on translational energy and product
vibrational states assigned from a spectroscopic, ro-vibrational
coupled energy expression based on the Dunham expansion is developed
and tested quantitatively. All models considered in this work
reproduce final state distributions determined from quasi-classical
trajectory (QCT) simulations with $R^2 \sim 0.98$. As a further
validation, thermal rates determined from the machine-learned models
agree with those from explicit QCT simulations and demonstrate that
the atomistic details are retained by the machine learning which makes
them suitable for applications in more coarse-grained
simulations. More generally, it is found that ML is suitable for
designing robust and accurate models from mixed
computational/experimental data which may also be of interest in other
areas of the physical sciences.
\end{abstract}

\section{Introduction}
Atom-diatom collisions at high collision energy are complex due to the
multitude of possible ways in which the available energy can be
redistributed. Relevant processes include reactions, exchange of
energy into translation, rotation, and vibration and depending on the
energy, low-lying electronic states. The number of possible states
that are accessible increases exponentially with the available
energy. To be able to understand bulk energy transfer in high-energy
processes, the probabilities of these processes need to be quantified
and understood.\\

\noindent
For a reactive atom-diatom collision, determining all state-to-state
cross sections is computationally very challenging even when treating
the process within classical mechanics. This is due to the large
number of reactant states $(\sim 10^4)$ that can combine with any
product state (also $\sim 10^4$) which gives rise to $\sim 10^8$
state-to-state cross sections. If such cross sections are determined
from quasi-classical trajectory (QCT) simulations, typically $10^4$ to
$10^5$ simulations are required for converged results which yields an
estimated $10^{12}$ to $10^{13}$ QCT simulations that would have to be
run at a given collision energy. Depending on the range of relevant
collision energies ($\sim 5$ eV in hypersonics) the number of required
QCT simulations can further increase by one or two orders of
magnitude. This is usually not possible nor desirable. With individual
QCT calculations taking roughly one second, even parallelization and
Moore's law scaling for improvement in processor speed will not
provide sufficient speedup for exhaustive calculations in the next
decades. Current computer technology limits tractable calculations to
$10^8$ meaning that any complete model for the probabilities
determined from QCT simulations would be supported by one in $\sim
10^7$ outcomes. In other words: either many final state distributions
are unconverged or the final states are not covered at all, or
both. This sparse representation creates a challenge for accurately
modeling the molecular dynamics in a way that is useful to larger
scale simulations. However, because state-to-state cross sections are
of paramount importance to determine rates for the process of
interest, alternative ways to address the problem are required as this
information is further used in reaction networks to model more
complicated chemistries.\\

\noindent
One application for which this is particularly relevant, is the high
temperature, high enthalpy flow prevalent in hypersonics. It is not
uncommon to find nominal temperatures that exceed 10000 K in shocks
and expansions that cause local non-equilibrium in the rovibrational
states of the chemically active molecules. Following the dynamics and
chemical development of rarefied gas flows employs primarily two
strategies. One uses computational fluid dynamics
(CFD)\cite{walpot:2012} which is based on the Navier Stokes
formulation of fluid dynamics and is valid for small Knudsen number
(ratio of the mean free path length to a physical length scale),
whereas direct simulation Monte Carlo (DSMC)\cite{dsmc:2017} is more
broadly applicable and is also valid for high Knudsen number. In DSMC,
particles (representing the physical atoms and molecules) move and
collide in physical space to which techniques from statistical
mechanics can be applied.\cite{boyd:2015} Particles in DSMC carry
information about their position, velocity, mass, size, and internal
state (for molecules) and move in cells within which they can collide
and exchange energy.\\

\noindent
DSMC cycles through the following steps: 1.) moving particles over a
time step $\Delta t$ smaller than the local mean free collision time;
2.) moving particles across cell boundaries or reflecting them at
solid boundaries; 3.) changing their internal states as a consequence
of collisions or reactions; 4.) sample average particle
information.\cite{boyd:2015} Step 3 is where microscopic information
about thermal reaction rates, state-to-state cross sections, and
vibrational relaxation rates enters. For chemical reactions most often
the ``total collision model'', based on a modified Arrhenius equation,
is used,\cite{dsmc} although more refined models are also
available.\cite{candler:1997} A more detailed and accurate description
is afforded by state-to-state or state-to-distribution
models. However, experimentally, it is very challenging or even
impossible to determine the relevant quantities at sufficiently high
temperature. Most cross sections presently used are derived from
chemical kinetics, many of which have not been measured at and above
3000 K, or can not be measured but are required at even higher
temperatures. Alternatively, the essential information can also be
obtained from quasi-classical trajectory (QCT) simulations using
state-of-the-art potential energy surfaces. Such an approach provides
all necessary information but implicitly assumes that the PESs and the
classical dynamics underlying QCT simulations are meaningful. In the
present work information from experimental spectroscopy is blended
into such a model.\\

\noindent
A widely used model for describing microscopic details of hypersonic
flow including chemical and relaxation processes is due to
Park.\cite{park:1993,park:2001} This approach considers separate
temperatures $T_{\rm v}$ and $T_{\rm t}$ for the vibration and
rotation/translation, respectively. In application to kinetics the
temperature is taken to be the geometric mean of these, the so-called
``$T-T_v$ model''. ``Chemistry'' enters such models through forward
and reverse reaction rates in the law of mass action for interacting
chemical species which are often determined from temperature-dependent
(modified) Arrhenius expressions. Following and extending the approach
from Millikan and White,\cite{millikan:1963} an important intuitive
correction established a framework for including vibrational
non-equilibrium in vibrational relaxation.\cite{park:1994}\\

\noindent
One problematic aspect of the Park model is that vibrational energy
becomes ``frozen'' above the translational energy because vibrational
relaxation is only included within the limits of Landau-Teller
theory\cite{landau:1936,millikan:1963,nikitin:2008} but the
contribution that accounts for removal of vibrational energy due to
dissociation of the product is neglected as is the movement of large
amounts of translational energy to vibrational energy via atom
exchange reactions. Including the contribution arising from
dissociation of the diatomic products was addressed and corrected in a
recent kinetic model which, however, neglected explicit coupling
between intramolecular rotation and
vibration.\cite{singh2020non,singh2020consistent} Also, with larger
computational platforms, it has been possible to investigate the
underlying physics on which the Park approach rests, namely the
two-temperature assumption and the preferential dissociation
model\cite{marrone:1963} which assumes that the amount of vibrational
energy removed during dissociation is large. For N$_2$ + N and N$_2$ +
N$_2$ it was shown that the $T-T_v$ model predicts a much faster N$_2$
dissociation for $T \leq 20000$ K than that obtained with direct
molecular simulation (DMS) whereas for $T = 30000$ K the two models
agree.\cite{candler:2016}\\

\noindent
Machine-learned (ML) models have proven to be effective in predicting
product distributions based on information about initial states
$(E_{\rm trans},v,j)$. For one, a neural network (NN) based
state-to-state (STS) model was conceived that predicts the cross
section for a given transition $(E_{\rm trans},v,j) \rightarrow
(E'_{\rm trans},v',j')$. This was demonstrated for the [NNO] reactive
system\cite{MM.sts:2019} in that not only the cross sections for
transitions that were not part of the training set were correctly
predicted, but also quantities derived from the cross sections -- such
as the total thermal rate $k(T)$ -- are in very good agreement with
those determined directly from QCT simulations. Conversely, a
distribution-to-distribution (DTD) model is capable of describing the
map between initial and final state distributions.\cite{MM.dtd:2020}\\

\noindent
From a practical perspective the most useful model is a
state-to-distribution (STD) model from which final state distributions
$P(E'_{\rm trans})$, $P(v')$, and $P(j')$ can be determined for every
initial state $(E_{\rm trans}, v, j)$. This is what is required for
more coarse-grained simulations, such as DSMC. The present work
presents state-to-distribution models for the N($^4$S)+O$_{2}$(X$^3
\Sigma_{\rm g}^{-}$) $\rightarrow$ NO(X$^2\Pi$) +O($^3$P) reaction
based on translational energy $E_{\rm trans}$ (instead of the diatom's
internal energy $E_{\rm int}$, see Ref.\cite{MM.std:2022}) and final
vibrational state assignment including mechanical ro-vibrational
coupling. It has been suspected earlier\cite{MM.std:2022} that
including such coupling may benefit model performance for high $(v,j)$
states. The trained ML models are based on data used for the earlier
STD model\cite{MM.std:2022} to allow for direct comparison between the
different approaches.\\

\noindent
The present work is structured as follows. First, the methods used are
presented. This is followed by an analysis of the data based on
product vibrational state assignments $v'$ using semiclassical
mechanics or from a model Hamiltonian (here a truncated Dunham
expansion). Then, the performance of NN-trained models based on
translational energy together with the two possibilities for assigning
final vibrational quantum numbers is assessed and compared. Finally,
thermal rates obtained via the two approaches are compared.\\

\section{Methods}
\subsection{Quasi-Classical Trajectory Simulations and Analysis}
The necessary data for ML-based models characterizing atom + diatom
collisions are based on quasi-classical trajectory (QCT)
simulations. The QCT data and an earlier STD model were used and
provided a means to compare the different
approaches.\cite{MM.std:2022} Additional QCT simulations were run as
needed using the same procedures as discussed
before.\cite{MM.std:2022} \\

\noindent
Initial conditions for QCT simulations were generated from
semiclassical quantization.\cite{kar65:3259,tru79} Such an approach
couples vibration and rotation in the sense that the initial
vibrational state is assigned from the numerical solution of an
integral involving the rotational barrier $j(j+1)/r^2$ where $r$ is
the diatomic bond length. After propagation of a specific initial
condition $(E_{\rm trans}, v, j)$ for a given impact parameter $b$,
the final states $(v',j')$ for the diatomic need to be determined and
the final translational energy $E_{\rm trans}'$ is obtained from the
relative velocities and reduced masses of the products. Assigning
$(v',j')$ quantum numbers is based on final momenta and positions
which are transformed to suitable coordinates. The internal final
angular momentum ${\bf j' = q' \times p'}$ for the product diatomic
species is determined from the final position ${\bf q'}$ and momentum
${\bf p'}$. Then the quadratic equation
\begin{equation}
\label{eq:rot}
    j' = -\frac{1}{2} +\frac{1}{2}\left(1+4\frac{{\bf j'}\cdot{\bf
        j'}}{\hbar^2}\right)^{\frac{1}{2}}
\end{equation}
is solved to determine the rotational quantum number ($j'$) as a
\textit{non-integer} number.\cite{kar65:3259,tru79} \\

\noindent
Using semiclassical (SC) mechanics, the non-integer vibrational
quantum number ($v'$) of the final diatomic species is calculated
according to\cite{kar65:3259,tru79}
\begin{equation}
    v_{\rm SC}' = -\frac{1}{2}
    +\frac{1}{\pi\hbar}\int^{r^+}_{r^-}\left\{2 \mu \left(E_{\rm int}
    -V(r)-\frac{{\bf j}\cdot{\bf
        j}}{2mr^2}\right)\right\}^{\frac{1}{2}}dr,
\end{equation}
where $r$ is the diatomic bond length, $r^+$ and $r^-$ are the turning
points of the diatomic species on the effective potential with
rotational state $j'$ for internal energy $E_{\rm int}'$, $\mu$ is the
reduced mass, and $V(r)$ is the potential energy curve of the product
diatom.\\

\noindent
\textit{Integer} ro-vibrational quantum numbers are then assigned as
the nearest integers $(v',j')$ using histogram binning. To ensure
conservation of total energy, the ro-vibrational energy $E_{\rm
  int}^{'}$ is recomputed from semiclassical
quantization\cite{kar65:3259,tru79} using the integer quantum numbers
$(v',j')$ and the final translational energy for the atom+diatom
system is adjusted according to $E_{\rm trans}^{'} = E_{\rm tot} -
E_{\rm int}^{'}$ where the final total internal energy is determined
from the final momenta and positions of the two atoms forming the
diatomic.\\

\noindent
For high $j'$ quantum numbers the angular momentum causes the
potential energy surface of the system to distort, changing the
characteristic frequency of the vibrations. A first-order
approximation to the energy based on the Dunham expansion is shown in
Eq. \ref{eq:coupledexp}. To include mechanical coupling between
vibration and rotation for the products, as afforded by a model
Hamiltonian (``MH''), the assignment of the final angular momentum
$j^{'}$ is retained as in Eq. \ref{eq:rot} but $v'$ is determined from
solving
\begin{equation}
\label{eq:coupledexp}
\begin{aligned}
E(v_{\rm MH}',j') = & \omega_e (v_{\rm MH}'+1/2) - \frac{\omega_e^2}{4
  D_e} (v_{\rm MH}'+1/2)^2 +B_e j'(j'+1) + D_e [j'(j'+1)]^2 \\& -
\alpha_e(v_{\rm MH}'+1/2)j'(j'+1)
\end{aligned}
\end{equation} 
for $v'$. Here, $E(v',j') = E_{\rm tot} - E_{\rm trans}'$,
$\omega_{e}$ is the harmonic frequency, $\omega_{e} x_{e}$ is the
first order correction, $B_e$ is the rotational constant, $D_e$ is the
centrifugal constant, and $\alpha_e$ is the ro-vibrational coupling
constant. For NO the data is $\omega_{e} = 1904.20$ cm$^{-1}$,
$\omega_{e} x_{e} = 14.08$ cm$^{-1}$, $B_e = 1.672$ cm$^{-1}$, $D_e =
0.00000054$ cm$^{-1}$, and $\alpha_e = 0.0171$
cm$^{-1}$.\cite{herzberg-I} The corresponding parameters on the
MRCI/aug-cc-pVTZ curve for NO are $\omega_e = 1871.0$ cm$^{-1}$ and
$\omega_e x_e = 14.04$ cm$^{-1}$.\\

\subsection{Neural Network}
One approach to creating a model that spans all of the possible
probabilities for the outcome of a collision between an atom and a
diatom is to use a neural network (NN) representation. Such an
approach was used previously in earlier STD work.\cite{MM.std:2022}
This NN consists of seven residual layers with two hidden layers per
residual layer, and uses 11 input and 254 output nodes corresponding
to the 11 features representing the initial reactant state and the 254
amplitudes characterizing the product state distributions. For
training, the NN inputs were standardized which ensures that the
distributions of the transformed inputs $x_{i}'$ over the entire
training data are each characterized by ($\bar{x}'_{i}=0$,
$\sigma'_{i}=1$). The NN outputs were normalized which ensures that
the distributions of the transformed outputs $x_{i}'$ over the entire
training data being characterized by ($\bar{x}'_{i}=\bar{x}_{i}$,
$\sigma'_{i}=1$). Standardization of the data generally yields a
faster convergence.\cite{lecun:2012} The loss function was the
root-mean-squared deviation between reference QCT data and model
predictions. As the NN outputs are probabilities, being non-negative
even after normalization, a softplus function is used as an activation
function of the output layer.\\

\noindent
The weights and biases of the NN were initialized according to the
Glorot scheme\cite{glorot2010understanding} and optimized using
Adam\cite{kingma2014adam} with an exponentially decaying learning
rate. The NN was trained using TensorFlow \cite{tf:2016} and the set
of weights and biases resulting in the smallest loss as evaluated on
the validation set were subsequently used for predictions. Overall,
final state distributions from 2184 initial conditions on a grid
defined by $(0.5 \leq E_{\rm trans} \leq 5.0)$ eV with $\Delta E_{\rm
  trans}=0.5$ eV $(5.0 \leq E_{\rm trans} \leq 8.0)$ eV with $\Delta
E_{\rm trans}=1.0$ eV, $v=[0,2,4,6,8,10,12,15,18,21,24,27,30,\\34,38]$,
and and $0 \leq j \leq 225$ with step size $\Delta j = 15$ were
generated, of which 7 were excluded due to low reaction
probability. This ``on-grid'' set of size $N=2184$, of which 7 data
points were excluded due to low reaction probability, was randomly
split into $N_{\rm train}=1700$ for training, $N_{\rm valid}=400$ for
validation, and $N_{\rm test}=77$ for testing.\cite{MM.std:2022} The
``on-grid'' set used for training, validation and test is
distinguished from ``off-grid'' data which refers to initial
conditions for which at least one of the entries in $(E_{\rm trans},
v, j)$ differs from an ``on-grid'' initial condition. All NNs in this
work were trained on a 3.60 GHz X 8 Intel Core i7-9700K CPU resulting
in training times shorter than 3 minutes. For additional technical
details, see Ref.\cite{MM.std:2022}\\

\section{Results}

\subsection{Data Preparation}  
A first important task in designing a ML model is the selection and
preparation of the data. Among other aspects, the present work
assesses whether a reliable and predictive STD model can be conceived
from training on / predicting of a) the translational energy $E_{\rm
  trans}'$ and b) the final vibrational quantum number $v_{\rm MH}'$
determined from a model Hamiltonian. The reason for this is that
$E_{\rm trans}'$ contains independent physical information from
$(v',j')$ states about the collision system whereas $E'_{\rm int}$ and
the $(v',j')$ states, as used in the previous model,\cite{MM.std:2022}
are somewhat redundant. Similarly, while assignment of $v'$ from
semiclassical quantization is one possibility, such an approach
neglects part of the mechanical $(v,j)$ coupling. Furthermore, using a
model Hamiltonian (such as a Dunham expansion\cite{dunham:1932} or a
Watson Hamiltonian,\cite{watson:1968} based on coefficients fit to
represent spectroscopically measured line positions or transitions)
for computing $v_{\rm MH}'$ includes valuable information from
experiment and makes the model somewhat less dependent on the level of
theory at which the intermolecular interactions have been / can be
determined in practice. This is the reason to explore changes in the
final vibrational distribution if a model Hamiltonian (here the Dunham
expression up to first order in coupling vibration and rotation) is
used. Retaining higher order terms is also possible to further refine
the approach but is not expected to fundamentally change the
findings.\\

\begin{figure}[h!]
\begin{center}
\includegraphics[width=0.95\textwidth]{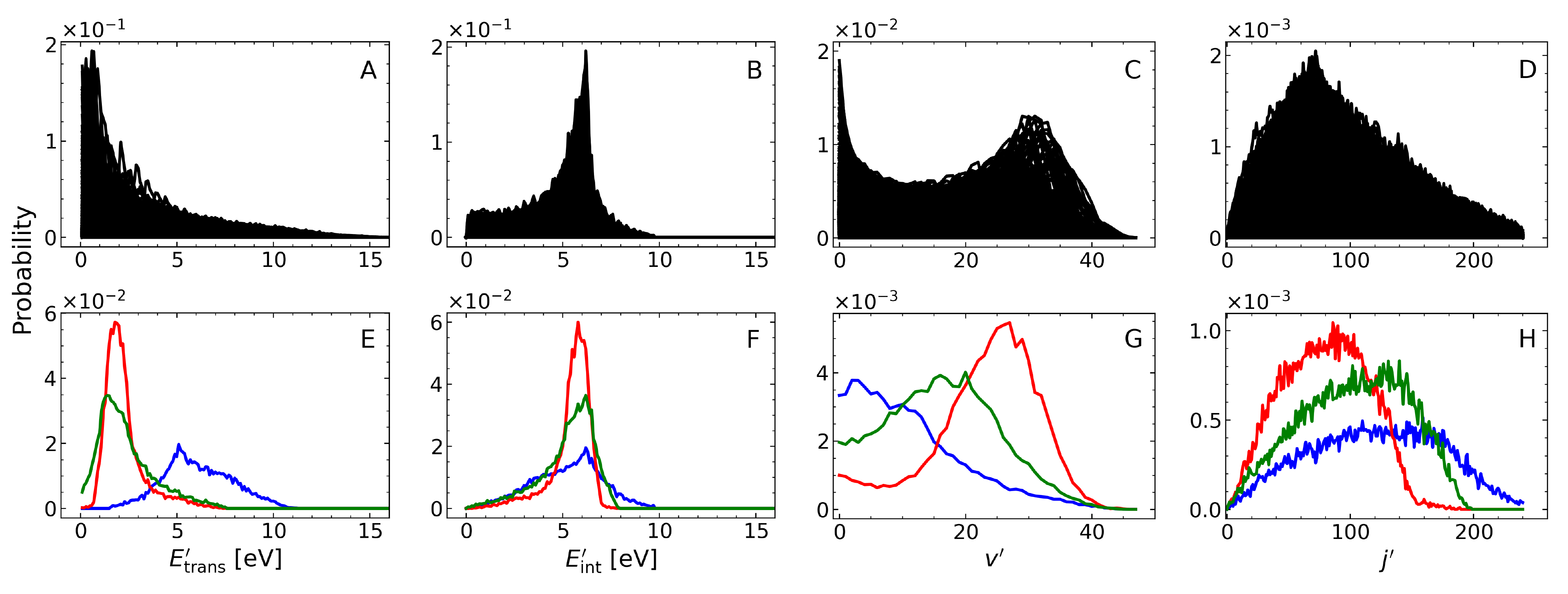}
\caption{Final state distributions $P(E_{\rm trans}')$ (panel A),
  $P(E_{\rm int}^{'})$ (panel B) $P(v^{'})$ (panel C) and $P(j^{'})$
  (panel D) from semiclassical quantization for NO$_2$ from QCT
  simulations for all 2184 initial reactant states. Panels E to H
  report selected distributions $P(E_{\rm trans}')$, $P(E_{\rm
    int}^{'})$, $P(v^{'})$, and $P(j^{'})$ to highlight their
  different shapes.}
\label{fig:fig1}
\end{center}
\end{figure}

\noindent
{\it $E'_{\rm int}$ versus $E_{\rm trans}'$:} The reason to employ
$E'_{\rm int}$ for training the original STD model\cite{MM.std:2022}
was that the number of grid points required to faithfully represent
$P(E'_{\rm int})$ was smaller than for $P(E_{\rm trans}')$ due to the
smaller span of energies and smoother features. Figures
\ref{fig:fig1}A and B report $P(E_{\rm trans}')$ and $P(E'_{\rm int})$
for all 2184 initial reactant conditions (``on-grid'') considered. It
appears that $P(E_{\rm trans}')$ extends to higher energies than
$P(E'_{\rm int})$. This is illustrated by considering a few select
final state distributions, see panels E and F in Figure
\ref{fig:fig1}.\\

\noindent
{\it Vibrational Quantum Number from Semiclassical Assignment and
  Ro-vibrational Energy Expression:} Final state distributions $P(v')$
and $P(j')$ from using semiclassical quantization in the final state
analysis are reported in Figure \ref{fig:fig1}. Panels C and D show
product state distributions corresponding to all 2184 initial reactant
states considered. For $P(v')$ they are found to extend out to $v'
\leq 45$ whereas for $P(j^{'})$ the highest final state is $j' \sim
240$ with maximum values $v'_{\rm max} = 47$, $j'_{\rm max} =
240$. Figures \ref{fig:fig1}G and H show individual final state
distributions and highlight the various shapes of these distributions
encountered depending on the initial condition.\\

\begin{figure}[h!]
\includegraphics[width=0.9\textwidth]{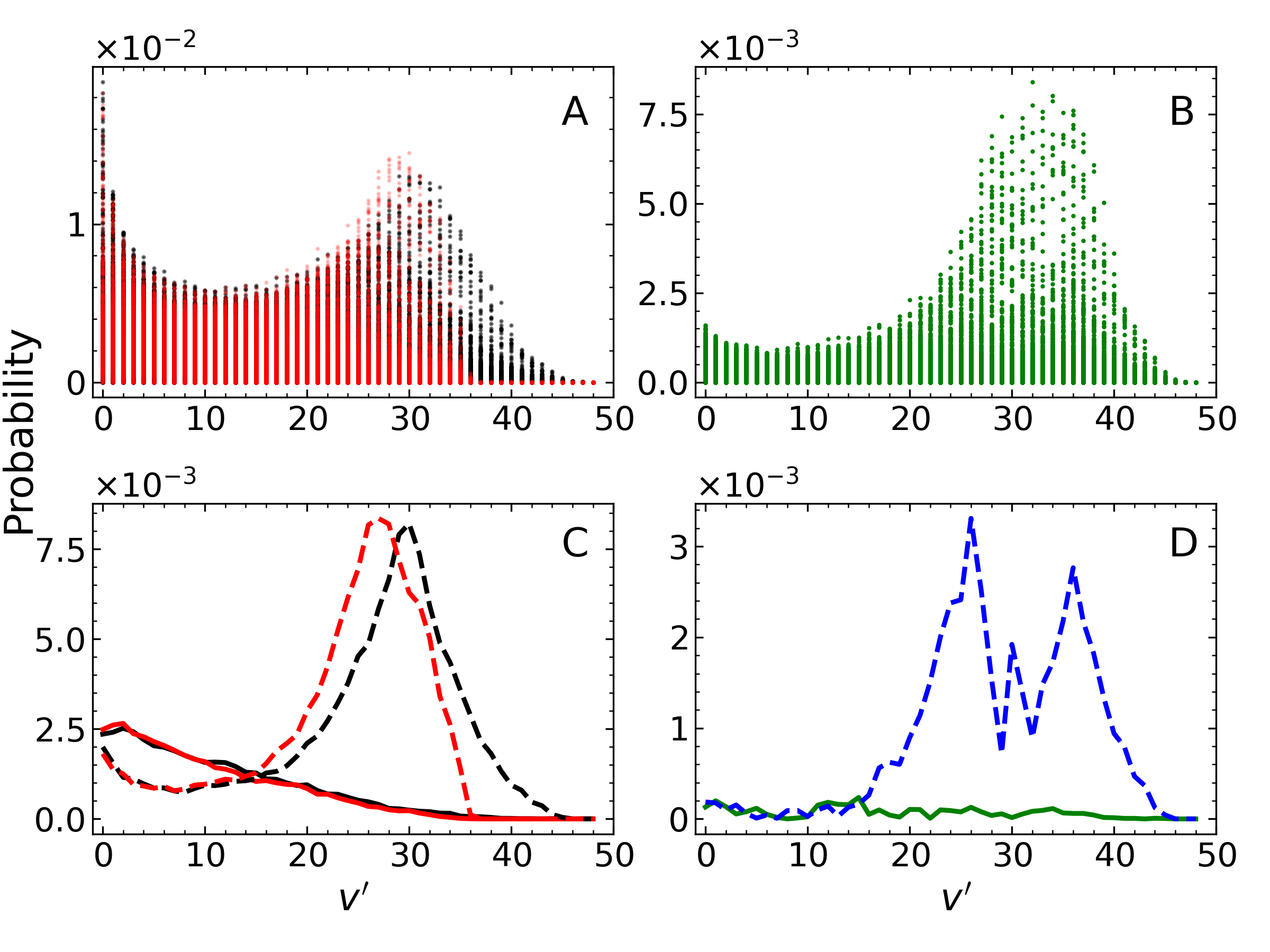}
\caption{Complete distribution for $P(v_{\rm SC}')$ (black) and
  $P(v_{\rm MH}')$ (red) for the set of 2184 initial conditions (Panel
  A) and for selected individual contributions (Panel C). The absolute
  difference between $P(v_{\rm SC}')$ and $P(v_{\rm MH}')$ is given in
  panels B (for complete set) $\&$ D (for selected individual
  distributions). Selected test initial conditions: $E_{\rm
    trans}=8.0, v=0, j=0$ (solid lines) and $E_{\rm trans}=1.0, v=30,
  j=0$ (dashed lines).}
\label{fig:fig2}
\end{figure}

\noindent
Next, the distributions from using SC and MH final states $v_{\rm
  SC}'$ and $v_{\rm MH}'$ are presented for all initial conditions,
see Figure \ref{fig:fig2}A. The black symbols denote $P(v_{\rm
  SC}')$ and extend up to $v_{\rm SC}' = 46$. This compares with a
final state distribution $P(v_{\rm MH}')$ from using the mechanically
coupled energy expression for the assignment of the final state (red
symbols) which only extends up to $v_{\rm MH}' = 36$. The difference
(green) between the two assignment schemes is reported in Figure
\ref{fig:fig2}B. For $v' \leq 20$ the difference in the
population is $\sim 10$ \% which increases to considerably higher
values for $v' \sim 30$ and above.\\

\noindent
Panels C and D of Figure \ref{fig:fig2} compare final state
distributions $P(v_{\rm SC}')$ and $P(v_{\rm MH}')$ for two specific
initial states. For $E_{\rm trans}=8.0, v=0, j=0$ (solid lines) the
two final vibrational state distributions are almost
indistinguishable, see also their absolute difference (green) in
Figure \ref{fig:fig2}D. On the other hand, for initial $E_{\rm
  trans}=1.0, v=30, j=0$ (dashed lines) the maximum for $P(v_{\rm
  SC}')$ (black) is at $v_{\rm SC}' = 30$ which shifts to $v_{\rm MH}'
= 26$ for $P(v_{\rm MH}')$. Evidently, the two distributions also
differ, see blue dashed line in Figure \ref{fig:fig2}D. The
general finding is that including mechanical ro-vibrational coupling
in assigning the final vibrational state $v_{\rm MH}'$ populates lower
states compared with an analysis based on a semiclassical approach.\\

\begin{figure}[h!]
\includegraphics[width=0.9\textwidth]{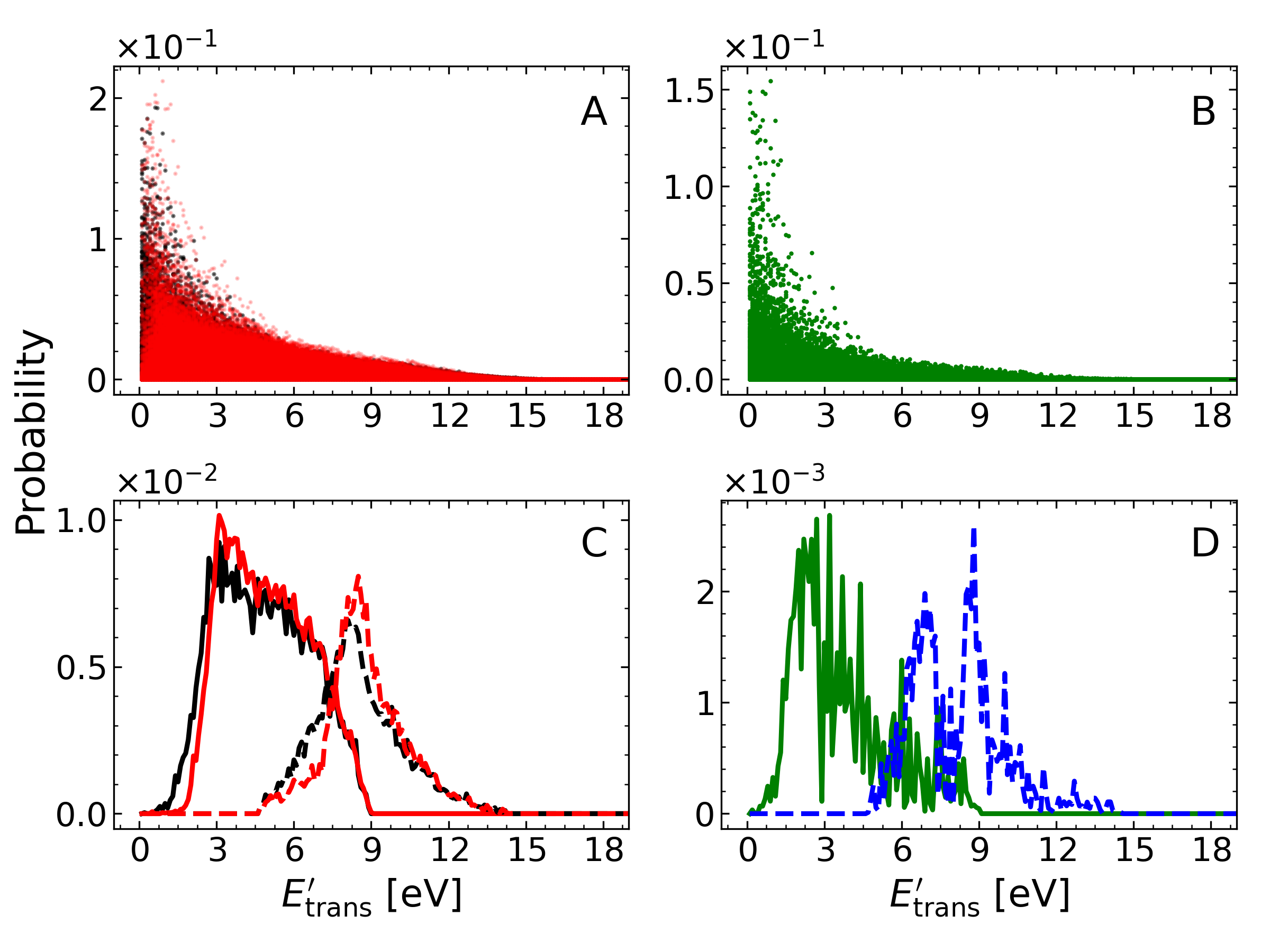}
\caption{Complete distribution for $P(E_{\rm trans, SC}')$ (black) and
  $P(E_{\rm trans, MH}')$ (red) for the set of 2184 initial conditions
  (Panel A) and for selected individual contributions (Panel C). The
  absolute difference between $P(E_{\rm trans, SC}')$ and $P(E_{\rm
    trans, MH}')$ is given in panels B (for complete set) $\&$ D (for
  selected individual distributions) for initial conditions $E_{\rm
    trans}=8.0, v=0, j=0$ (solid lines) and $E_{\rm trans}=8.0, v=24,
  j=120$ (dashed lines).}
  \label{fig:fig3}
\end{figure}

\noindent
Assignment of the final state to integer values $(v_{\rm MH}',j')$
leads to differences in the corresponding internal energy $E_{\rm
  int}'(v_{\rm MH}',j')$ from the value obtained when considering
$E_{\rm int}' = E_{\rm tot}-E'_{\rm trans}$ based on energy
conservation. Hence, the difference $E_{\rm int}'(v_{\rm MH}',j') -
E_{\rm int}'$ has to be redistributed into $E'_{\rm trans}$ which
leads to $E_{\rm trans, MH}'$ from including mechanical coupling in
the rovibrational energy. Figure \ref{fig:fig3}A compares the final
translational energy distributions $P(E_{\rm trans, SC}')$ (black) and
$P(E_{\rm trans, MH}')$ (red). The overall shapes of the two
distributions are comparable but when considering the absolute
difference between the two distributions (Figure \ref{fig:fig3}B,
green) variations up to 50 \% are found. For high translational energy
the absolute differences decay to zero.\\

\noindent
Focussing on two specific initial conditions ($E_{\rm trans}=8.0, v=0,
j=0$ and $E_{\rm trans}=8.0, v=24, j=120$, see Figure \ref{fig:fig3}C)
indicate that the overall shapes of the final translational energy
distributions depend little on whether mechanical coupling was
included or not to determine $v'$. For initial ($E_{\rm trans}=8.0,
v=0, j=0$) nonzero probability at low translational energy starts at a
higher value for an analysis based on MH (red) compared with SC
(black). A similar behaviour is seen for the initial condition given
by ($E_{\rm trans}=8.0, v=24, j=120$) (dashed lines in Figure
\ref{fig:fig3}C). Absolute differences are reported in Figure
\ref{fig:fig3}D and reach a maximum of $\sim 20$ \%.\\

\subsection{Trained STD Models}
Next, full STD models were trained based on assignment of the final
vibrational state from semiclassical analysis ($v_{\rm SC}'$) or using
a first-order Dunham expansion ($v_{\rm MH}'$) together with $E_{\rm
  trans}$, respectively. In the following, ``on-grid'' values refer to
initial conditions that were used for training the NN (training,
validation, test data - see Methods) and ``off-grid'' corresponds to
initial conditions that differed from ``on-grid'' values in at least
one of the initial conditions $(E_{\rm trans}, v, j)$.\\

\begin{figure}[h!]
\begin{center}
\includegraphics[width=0.99\textwidth]{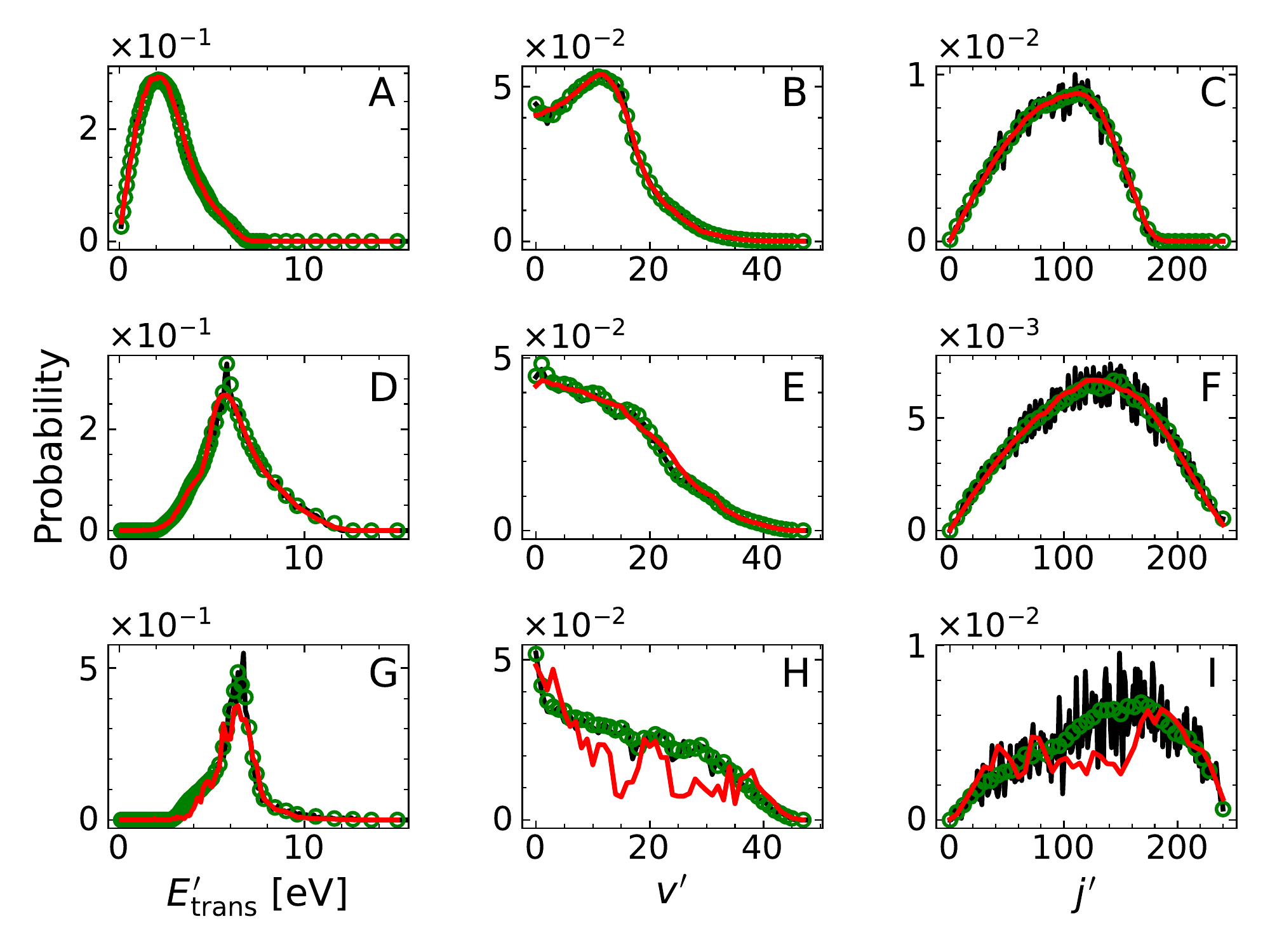}
\caption{Comparison of product state distributions from QCT (symbols)
  and STD model evaluation (red solid line) for SC, for initial
  conditions from the ``on-grid'' set (excluding $E_{\rm trans}=0.5$
  eV). STD predictions are for the initial condition for which the
  prediction is best (top, largest $R^2$), is closest to the average
  $R^2$ (middle), and worst (bottom, lowest $R^2$), respectively.  The
  corresponding $R^2$ values for ($P(E'_{\rm trans})$, $P(v')$,
  $P(j')$) are: [0.9985,0.9985,0.9995] (best); [0.9880,0.9954,0.9895]
  (closest to mean) and [0.9307,0.4082,0.3578] (worst). The
  corresponding reaction probabilities $P_r =N_{r}/N_{\rm tot}$ are
  0.378, 0.217, and 0.064 for the best, average and worst prediction,
  respectively. Here, $N_{\rm tot}$ and $N_r$ the total number and
  number of reactive trajectories, respectively.}
\label{fig:fig4}
\end{center}
\end{figure}

\noindent
The performance of the trained models in terms of $R^2$ and RMSD is
summarized in Tables \ref{tab:tab1} and \ref{sitab:tab1}. The analysis
is carried out for all on-grid-values, excluding seven distributions
with low probabilities ($P(v') < 0.005)$), and excluding the lowest
translational energy $E_{\rm trans}=0.5$ eV because the number of QCT
simulations may be insufficient to fully converge these final state
distributions. In addition, this analysis was also carried out for
initial conditions off-grid in $(v,j)$ and at fixed $E_{\rm trans} =
4.0$ eV. For all degrees of freedom and all test sets $R^2 > 0.97$
which indicates reliable statistical models. Such $R^2$ measures are
also comparable to the performance if instead of using $E_{\rm trans}$
the models are trained on $E_{\rm int}$ for which $R^2 > 0.98$ was
found.\cite{MM.std:2022} If simulations at the lowest translational
energy (0.5 eV) are excluded, the performance is somewhat improved for
all degrees of freedom. Finally, whether the final vibrational state
$v'$ was assigned from a semiclassical treatment or a mechanically
coupled energy expression has negligible influence on the performance
of the models. When using the RMSD as the measure to compare reference
distributions with predictions from the trained STD models (see Table
\ref{sitab:tab1}) similar conclusions are drawn as for $R^2$.\\

\begin{table}[h]
\begin{center}
\caption{$R^2$ between QCT results and the trained STD model based on
  $E_{\rm trans}$ and either semiclassical or mechanically coupled
  determination of $v'$. On-grid (All) is for all initial conditions
  from the training, validation and test set for all translational
  energies considered. The numbers in bracket are from the test set
  (77 initial conditions) only. On-grid* and Off-grid* are for initial
  conditions with $E_{\rm trans} = 4.0$ eV only and Off-grid* contains
  all initial conditions for which either $v$, $j$, or both $(v,j)$
  differed from the values used for the training, validation and test
  sets.}
    \begin{tabular}{lcc|cc}
    \hline
        \hline
      & On-grid (All)  & On-grid (test) & On-grid* &  Off-grid* \\
    \hline
    $P(E_{\rm trans,SC}')$   & 0.9878 & 0.9844  & 0.9949  & 0.9918   \\
    $P(E_{\rm trans,MH}')$     & 0.9737 & 0.9840 & 0.9959  & 0.9942 \\
    \hline
    $P(v_{\rm SC}')$   & 0.9881 & 0.9835  & 0.9949  & 0.9961   \\
    $P(v_{\rm MH}')$     & 0.9834 & 0.9817 & 0.9966  & 0.9972 \\
    \hline
    $P(j_{\rm SC}')$   & 0.9845 & 0.9816  & 0.9940  & 0.9950   \\
    $P(j_{\rm MH}')$     & 0.9879 & 0.9882 & 0.9950  & 0.9951 \\
    \hline
    \hline
    ``SC'' (overall)   & 0.9868 & 0.9832  & 0.9946  & 0.9943   \\
    ``MH'' (overall)     & 0.9817 & 0.9846 & 0.9959  & 0.9955 \\
    \hline
    \hline
  \end{tabular}
  \label{tab:tab1}
  \end{center}
\end{table}

\begin{figure}[h!]
\begin{center}
\includegraphics[width=0.9\textwidth]{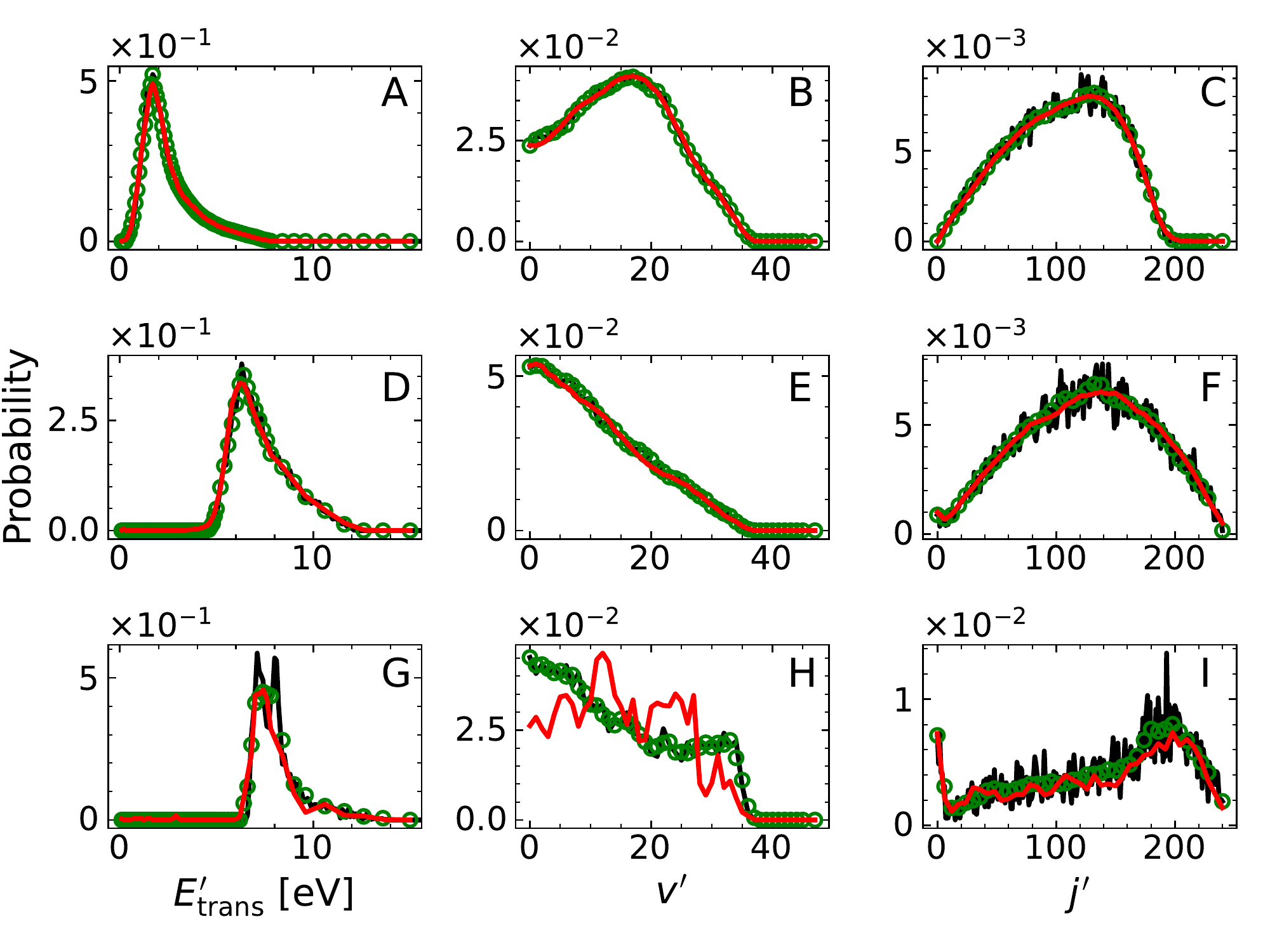}
\caption{Comparison of product state distributions from QCT (symbols)
  and STD model evaluation (red solid line) for MH, for initial
  conditions from the ``on-grid'' set (excluding $E_{\rm trans}=0.5$
  eV). STD predictions are for the initial condition for which the
  prediction is best (top, largest $R^2$), is closest to the average
  $R^2$ (middle), and worst (bottom, lowest $R^2$), respectively. The
  corresponding $R^2$ for ($P(E'_{\rm trans})$, $P(v')$, $P(j')$) are:
  [0.9987,0.9987,0.9991](best); [0.9911,0.9971,0.9911](closest to
  mean) and [0.9741,0.5043,0.8305](worst).  The corresponding reaction
  probabilities $P_r =N_{r}/N_{\rm tot}$ are 0.365, 0.308, and 0.044
  for the best, average and worst prediction, respectively. Here,
  $N_{\rm tot}$ and $N_r$ the total number and number of reactive
  trajectories, respectively.}
\label{fig:fig5}
\end{center}
\end{figure}

\noindent
An explicit comparison of final state distributions from models based
on SC (Figure \ref{fig:fig4}) and MH (Figure \ref{fig:fig5})
for assigning the final vibrational state is provided for the
best-performing, for an average-performing, and for the least
performing distribution. The QCT final state distributions are shown
as green open circle and the STD predictions as the red solid
line. For the best-performing prediction the STD model closely follows
the target data. This is still the case for a prediction that
represents an average performance. It is also notable that the shape
of the distributions for the same degree of freedom can change
appreciably, depending on the initial condition. For the
worst-performing prediction it is found that SC (see Figure
\ref{fig:fig4}) leads to a visually inferior model compared to
using MH (Figure \ref{fig:fig5}). In both worst-performing cases,
the number of reactive trajectories $N_r$ as a fraction of the total
number of trajectories $N_{\rm tot}$ is only $\sim 5$ \% and
increasing the sampling may also lead to better reference data.\\

\begin{figure}[h!]
\includegraphics[width=0.9\textwidth]{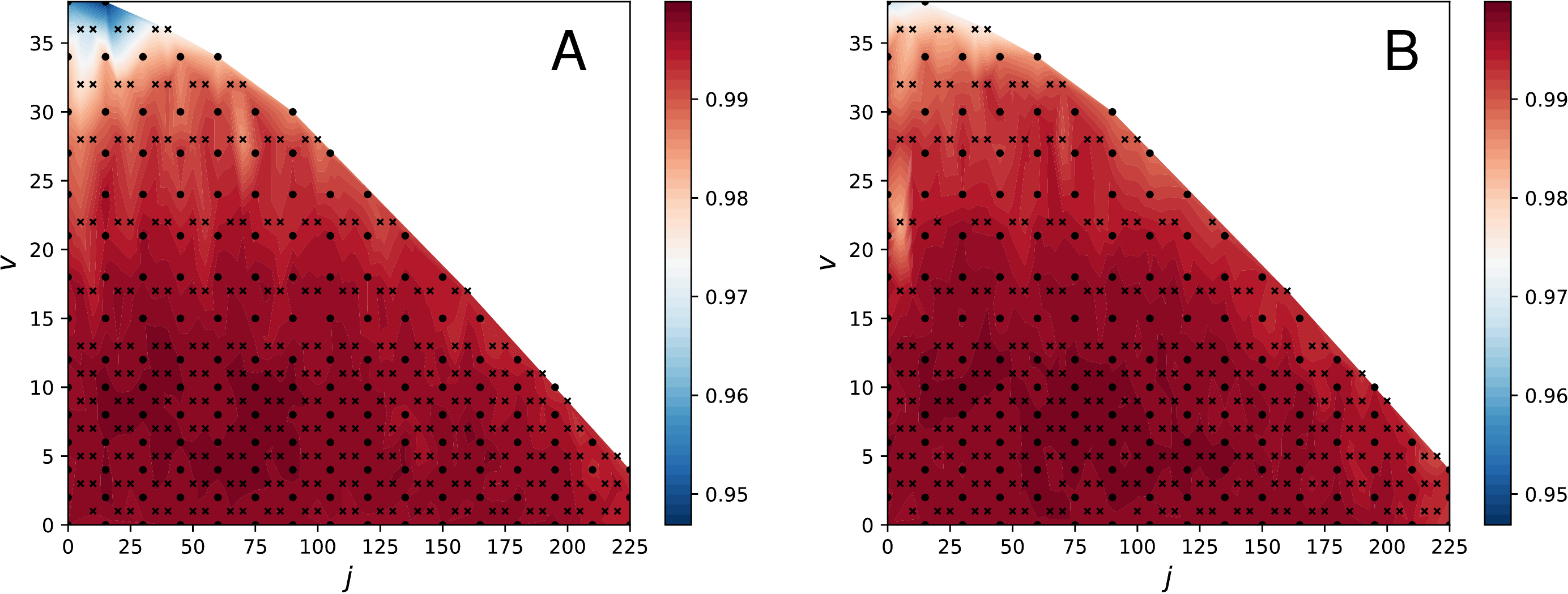}
\caption{2D map of the averaged (over 5 independently trained STD
  models) $R^{2}$ values between QCT data ($P(E'_{\rm trans}), P(v'),
  P(j')$) and STD evaluation for initial conditions ``on-grid'' (see
  Methods, circles) and ``off-grid''
  ($[v=1,3,5,7,9,11,13,17,22,28,32,36],\\ j=[10, 20, 25, 35, 40, 50,
    55, 65, 70, 80, 85, 100, 110, 115, 125, 130, 140, 145, 155, 160,
    170, 175,\\ 185, 200,205, 215, 220]$, crosses). Panel A for ``SC''
  and panel B for ``MH'' assignment of the final vibrational
  state. All QCT simulations were carried out at $E_{\rm trans}= 4.0$
  eV.}
\label{fig:fig6}
\end{figure}

\noindent
For a more global characterization of the final states for the entire
range of possible initial states $(v,j)$, 5 independent models were
trained from reference data for SC (Figure \ref{fig:fig6}A) and MH
(Figure \ref{fig:fig6}B). The averaged $R^2$ over all 5 models using
``on-grid'' (see Methods; black circles) and ``off-grid'' (see caption
Figure \ref{fig:fig6}; crosses) data for each of the initial states
considered shows a uniformly high performance for both approaches with
typical $R^2 \sim 0.98$ or better. For the highest $v-$values the
absolute difference between using MH and SC expressions for $v'$
becomes most apparent. There, using a ``MH'' model appears to slightly
outperform an analysis based on a ``SC'' energy expression. But the
differences are insignificant.\\

\subsection{Thermal Rates}
\begin{figure}[h!]
\begin{center}
\includegraphics[width=0.6\textwidth]{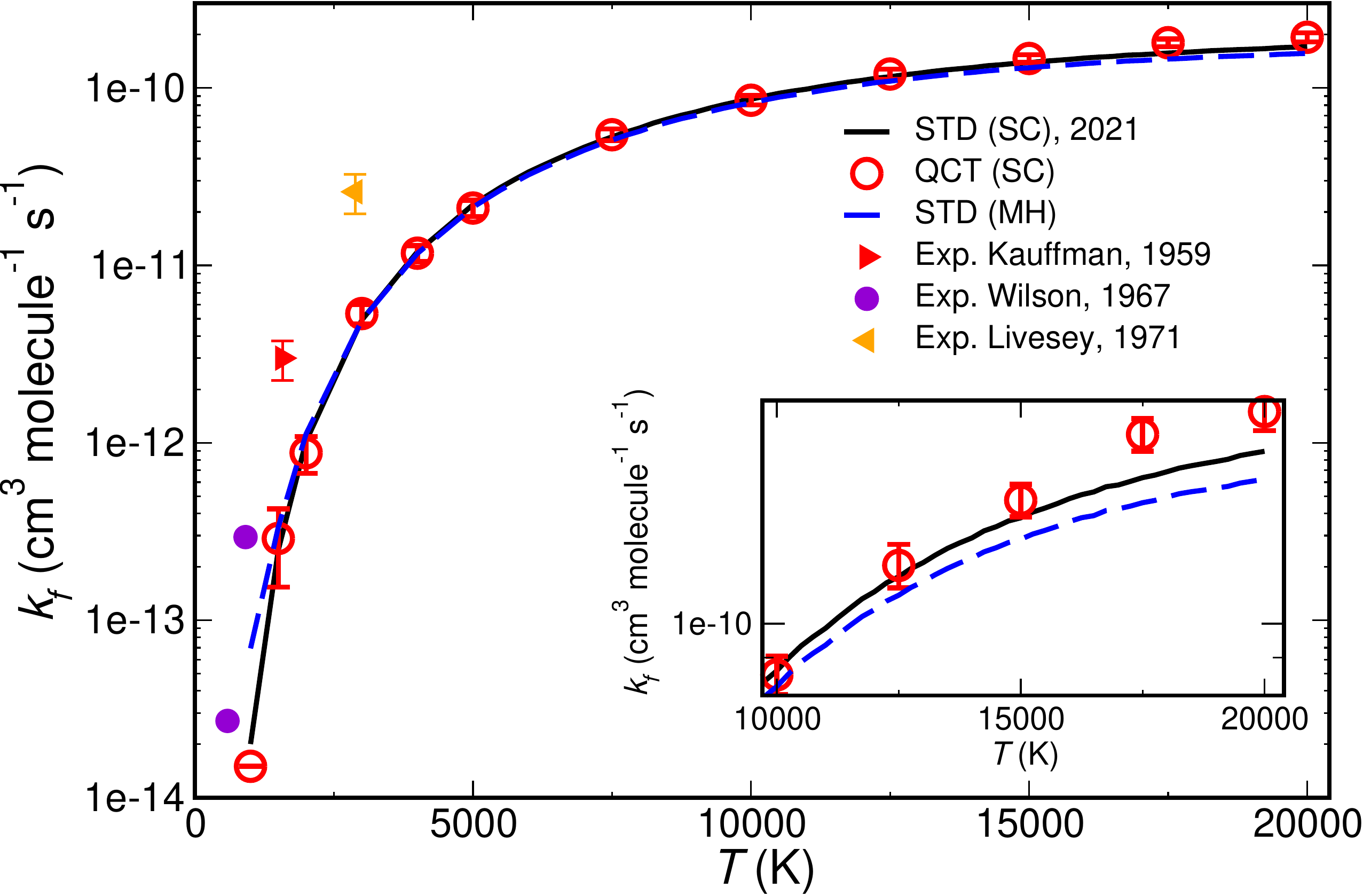}
\caption{The thermal forward rate $k_{f}$ calculated from QCT (open
  red circle) and STD model (solid black line) for the $^{4}$A$^{'}$
  state of the N($^{4}$S) + O$_2$(X$^{3} \Sigma_{g}^{-}$ )
  $\rightarrow$ NO(X$^2 \Pi$) + O($^3$P) reaction between 1000 and
  20000 K.  Additional simulations were made with the STD model based
  on $E_{\rm trans}$ and ``MH'' (blue dashed line). The present rates
  agree quantitatively with those directly obtained from QCT
  simulations, in particular considering the reduction of required
  compute time by 7 orders of magnitude or more. Rates for ``MH'' are
  expected to change slightly if higher-order terms in the Dunham
  expansion are included. Experimental total forward reaction rate
  $k_{\rm f}$ (including contributions from the doublet and the
  quartet states) are also shown for comparison: (red
  triangle)\cite{Kaufman:1959}, (orange triangle)\cite{Livesey:1971}
  and (magenta circle)\cite{wilson1967rate}.}
\label{fig:fig7}
\end{center}
\end{figure}

It is also of interest to determine thermal rates from the trained
models. In general, such a rate is determined from the reaction
probability $P_r$ according to
\begin{equation}
    k(T)=g(T) \sqrt{\frac{8k_{\rm B}T}{\pi \mu}}\pi b^{2}_{\rm max} P_r.
\end{equation}
For QCT simulations based on stratified sampling of the impact
parameter $b$, $P_{r}=\sum_{k=1}^{K} V_{k} \frac{N_{k}^{r}}{N^{\rm
    tot}_{k}}$ where $N^{r}_{k}$ and $N^{\rm tot}_{k}$ are the number
of reactive and total trajectories, respectively, for stratum $k$ with
impact parameter $b_k$. The fractional volumes $V_k = \frac{b_k^2 -
  b_{k-1}^2}{b_{\rm max}^2}$ of stratum $k$ obey $\sum_{k=1}^K V_k =
1$. For the STD model the reaction probability can be obtained as $P_r
= \int_{E=0}^{E_{\rm max}} P(E)dE$ where $E = E_{\rm trans}'$ and
$E_{\rm max} = 15.0$ eV. For the forward N($^4$S) + O$_{2}$(X$^3
\Sigma_{\rm g}^{-}$) $\rightarrow$ NO(X$^2\Pi$) + O($^3$P) reaction in
the $^4$A$'$ electronic state the degeneracy factor $g(T) = 1/3$ and
$\mu$ is the reduced mass of the reactants.\cite{MM.no2:2020} The two
approaches are compared in Figure \ref{fig:fig7} and favourable
agreement is found over a wide temperature range. Hence, the STD model
can also be used to determine macroscopic quantities such as realistic
reaction rates which is essential.\\

\section{Discussion and Conclusion}
The present work discusses a rapid, accurate, and physics-based model
to determine final state distributions for given initial conditions
$(E_{\rm trans}, v, j)$ for atom+diatom reactions. It is shown that
training on $E_{\rm trans}'$ and $v'$ from an energy expression
including mechanical coupling (Dunham expression) yields a model
performance of $R^2 \sim 0.98$ or better compared with rigorous QCT
reference simulations. Models based on $(E_{\rm trans}, v, j)$ are
more physically meaningful than those using $(E_{\rm int}, v, j)$
because $E_{\rm int}$ and $(v,j)$ are partly redundant whereas $E_{\rm
  trans}$ contains new, complementary, and physically relevant
information. The gain in speed compared to explicit QCT simulations is
about 7 orders of magnitude required for inference (i.e., excluding
training the NN which is minutes, and generating the required QCT
reference data for training) assuming that $10^5$ QCT trajectories are
sufficient for converged final state distributions. Hence, this is a
promising avenue to be used in more coarse-grained simulations of
reaction networks relevant to hypersonics and combustion.\\

\begin{figure}[h!]
\begin{center}
\includegraphics[width=0.9\textwidth]{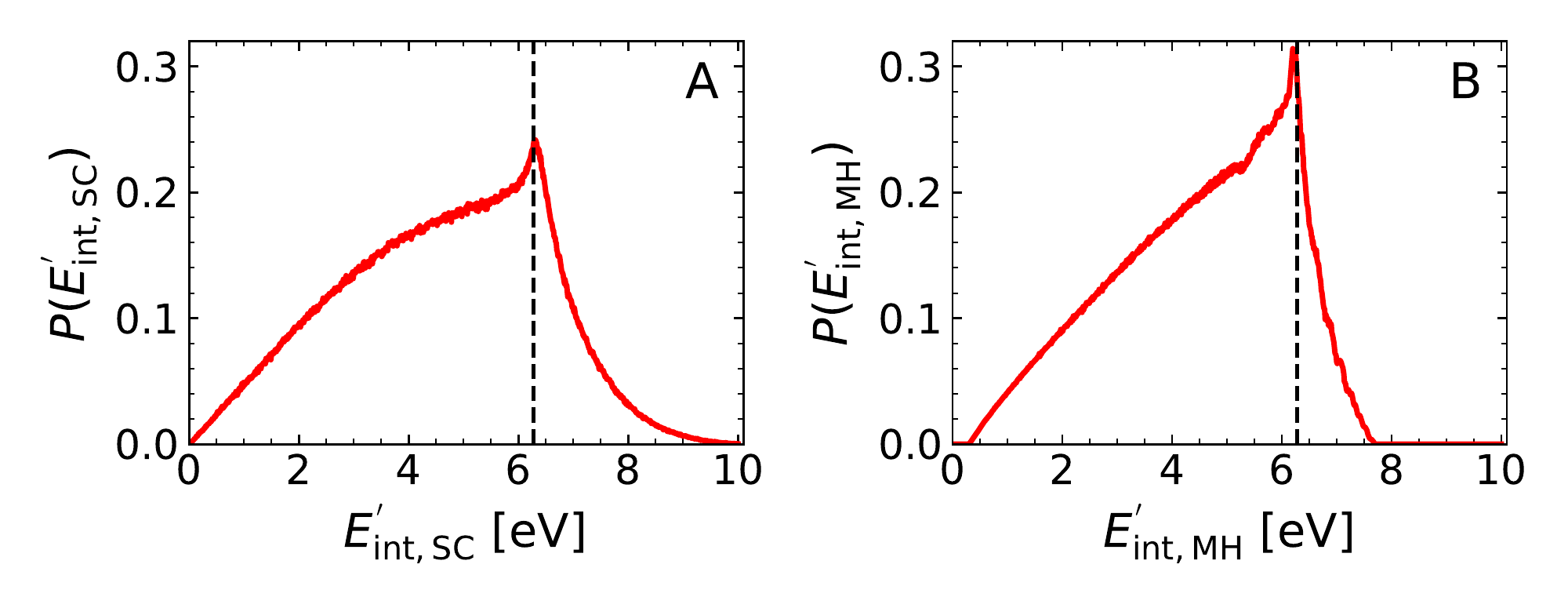}
\caption{Final internal energy distribution $P(E_{\rm int}')$
  depending on the assignment scheme (SC, MH) used. Distributions are
  generated from the complete data set (on-grid and off-grid) used as
  initial conditions in the QCT simulations. The dissociation energy
  for NO (dashed black) is shown as well (6.27 eV and 6.29 eV for SC
  and MH, respectively, compared with 6.50 eV from
  experiment.\cite{herzberg-I}) and the percentage of population above
  dissociation is 21 \% and 14 \% for SC and MH, respectively. The
  $D_e$ value for MH was determined from using a scaling factor
  between experimental $D_e$, and $D_e = \omega_e^2 / (4 \omega_e
  x_e)$ using spectroscopic constants.\cite{herzberg-I}}
\label{fig:fig8}
\end{center}
\end{figure}

\noindent
Including coupling in determining the final $v'$ state of the diatomic
by means of a truncated Dunham expansion as a typical model
Hamiltonian (MH) leads to population of lower $v'$ states than
assignment from semiclassical mechanics (SC). The overall shapes of
the final state distributions $P(v'_{\rm SC})$ and $P(v'_{\rm MH})$ do
not change appreciably, see for example Figure
\ref{fig:fig2}. However, final vibrational distributions from SC
extend to higher values $v'$ than those from using a Dunham
expansion. The influence of this effect is illustrated in Figure
\ref{fig:fig8} which compares final internal energy distributions from
the two assignment schemes (SC and MH) for all on-grid initial
conditions. The relevant NO dissociation energies from the SC and MH
models are also indicated and allow to determine the fraction of final
states that are above this threshold to be 21 \% and 14 \%,
respectively, for SC and MH.\\

\noindent
If larger products than diatomics are produced in such reactions, it
is expected that the MH approach for assigning final vibrational
states is more convenient than an assignment based on semiclassical
mechanics. Furthermore, using effective Hamiltonians to determine
internal final states of molecular fragments opens the possibility to
blend accurate spectroscopic information into such models which makes
them less dependent on the level of quantum chemical theory at which
the intermolecular interactions can be practically described. For
example, multi-reference configuration interaction (MRCI) calculations
with extended basis sets (triple zeta and larger) for molecules with
more than 3 atoms can quickly become computationally
prohibitive. However, it should also be noted that both schemes have
their advantages and drawbacks. For experimental observables, such as
thermal rates, both models perform on par, see Figure
\ref{fig:fig7}.\\

\noindent
The actual population of highly excited states in the product diatom
is particularly relevant for hypersonics because such states can
easily dissociate during the next collision and the ensuing products
(free atoms at high translational energy) induce rich chemistry. In
general, highly vibrationally excited states are correlated with
facile dissociation whereas high $j-$values prevent and delay breakup
of the diatomic. Furthermore, the vibrational coordinate can also lead
to electronic state relaxation rate differences in subsequent
reactions.\cite{boyd:2016} Hence, overall, depending on the assignment
scheme used for $(v,j)$ the energy partitioning into other degrees of
freedom (translation, electronic) will be affected, too.\\

\noindent
In summary, the present work demonstrates that it is possible to
construct machine-learned models for final state distributions given
state-specific information for the reactants for atom + diatom
reactions. The trained models are accurate, rapid to evaluate, and can
be extended as needed for other applications, e.g. combustion, by
supplying suitable training data. It is anticipated that such
approaches are beneficial to perform more coarse-grained modeling to
applications in hypersonic flow and computational combustion studies.

\section*{Data and Code Availability}
Exemplary data sets and code for training and evaluating STD models
can be found at \url{https://github.com/MMunibas/STDMH}.

\section*{Supporting Information}
The supporting information contains one additional Table.

\section{Acknowledgments}
This work was supported by AFOSR, the Swiss National Science
Foundation through grants 200021-117810, 200020-188724, the NCCR MUST,
and the University of Basel (to MM). J.A. acknowledges financial
support from the Swiss National Science Foundation individual grant
(Grant No. 200020$_{-}$200481). Discussions with Dr. D. Koner are
greatly appreciated.

\bibliography{refs.clean}

\newpage

\section{Table of Contents Graphics}
\begin{figure}[H]
  \centering \includegraphics[width=0.50\linewidth]{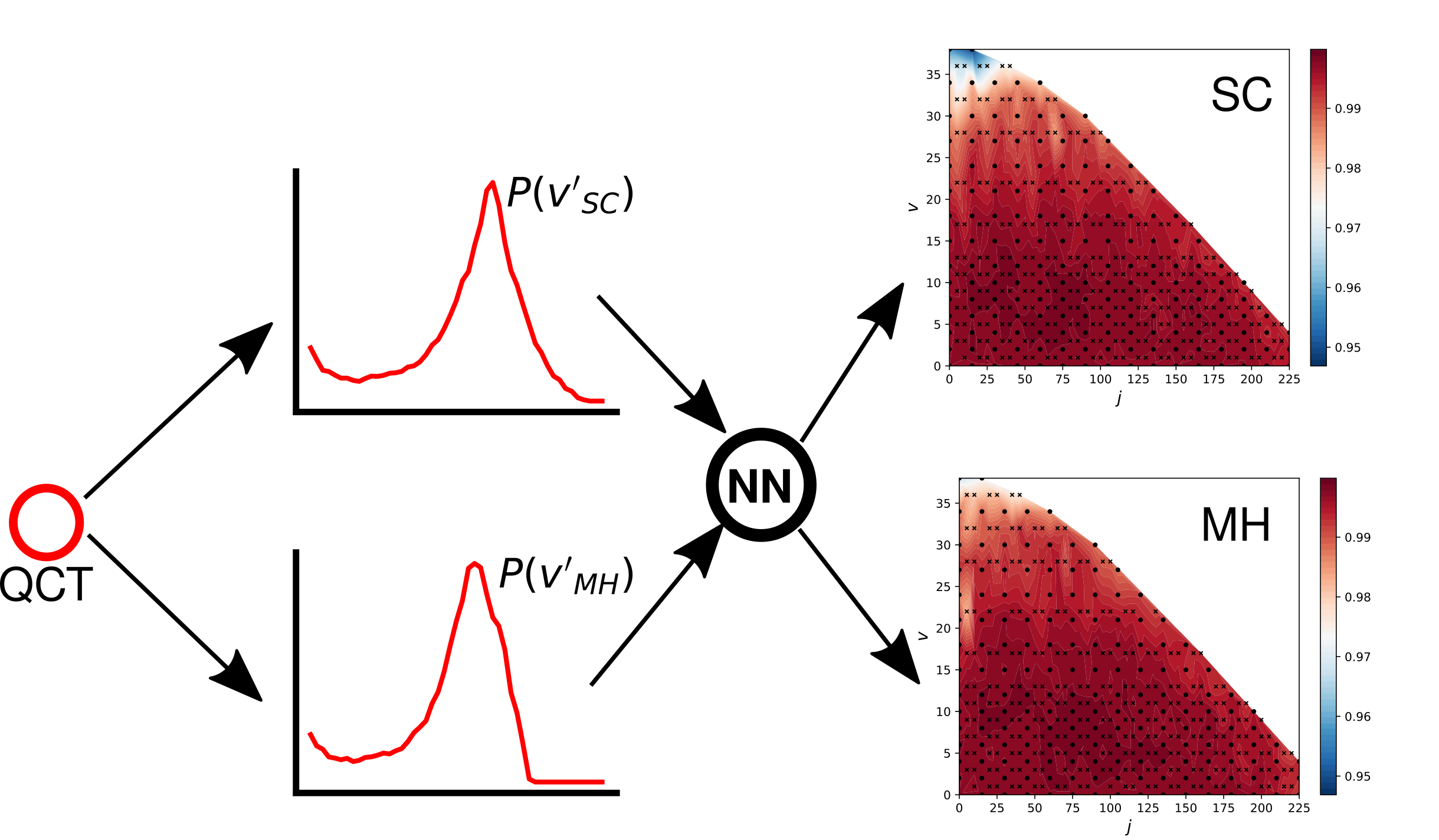}
    \label{fig:toc}
\end{figure}

\newpage
\clearpage

\section{Supporting Information:}

\maketitle

\begin{table}[h]
\begin{center}
  \caption{RMSD between QCT results and the trained STD model based on
    $E_{\rm trans}$ and either semiclassical or mechanically coupled
    determination of $v'$. On-grid (All) is for all initial conditions
    from the training, validation and test set for all translational
    energies considered. The numbers in bracket are from the test set
    (77 initial conditions) only. On-grid* and Off-grid* are for
    initial conditions with $E_{\rm trans} = 4.0$ eV only and
    Off-grid* contains all initial conditions for which either $v$,
    $j$, or both $(v,j)$ differed from the values used for the
    training, validation and test sets.}
    \begin{tabular}{lcc|cc}
    \hline
        \hline
      & On-grid (All) & On-grid (test)  & On-grid* &  Off-grid* \\
    \hline
    $P(E_{\rm trans, SC}')$   & 0.0110 & 0.0134  & 0.0053 & 0.0080   \\
    $P(E_{\rm trans, MH}')$     & 0.0143 & 0.0151 & 0.0056 & 0.0081 \\
    \hline
    $P(v_{\rm SC}')$   & 0.0015 & 0.0019  & 0.0011  & 0.0011   \\
    $P(v_{\rm MH}')$     & 0.0019 & 0.0025 & 0.0011  & 0.0010 \\
    \hline
    $P(j_{\rm SC}')$    & 0.0002 & 0.0003  & 0.0002  & 0.0002   \\
    $P(j_{\rm MH}')$     & 0.0002 & 0.0003 & 0.0002  & 0.0002 \\
    \hline
    \hline
    ``SC'' (overall)   & 0.0042 & 0.0052  & 0.0022  & 0.0031   \\
    ``MH'' (overall)    & 0.0055 & 0.0059  & 0.0028  & 0.0031 \\
    \hline
    \hline
  \end{tabular}
  \label{sitab:tab1}
  \end{center}
\end{table}

\end{document}